\RequirePackage[2020-02-02]{latexrelease}

\documentclass[aip, jap, twocolumn, superscriptaddress, citeautoscript]{revtex4}

\setcitestyle{super}

\usepackage{amsfonts, amsmath, bm, graphicx}
\usepackage{epstopdf}

\newcommand{\rand}{\ensuremath{\,\textrm{rand}}}
\newcommand{\sgn}{\ensuremath{\,\textrm{sgn}}}

\begin{document}

\title{Stochastic generation in a Josephson-like antiferromagnetic spin Hall oscillator driven by a pure AC current}

\author{D.~V.~Slobodianiuk} \email[Corresponding author, e-mail: ]{denslobod@ukr.net}
\affiliation{Taras Shevchenko National University of Kyiv, 64/13 Volodymyrs'ka str., 01601, Kyiv, Ukraine}
\affiliation{Institute of Magnetism of the NAS of Ukraine and the MES of Ukraine, 36-b Acad.~Vernadskogo blvd., 03142, Kyiv, Ukraine}
\author{O.~V.~Prokopenko}
\affiliation{Taras Shevchenko National University of Kyiv, 64/13 Volodymyrs'ka str., 01601, Kyiv, Ukraine}

\date\today


\begin{abstract}
We demonstrate numerically that a pure time-harmonic bias AC current of some particular amplitude $\tau_f$ and angular frequency $\omega_f$ can excite the chaotic magnetization dynamics in a Josephson-like antiferromagnetic (AFM) spin Hall oscillator (SHO) having a biaxial magnetic anisotropy of an AFM layer.
The nature of such a stochastic generation regime in a Josephson-like AFM SHO could be explained by the random hopping of the SHO's work point between several quasi-stable states under the action of applied AC current.
We revealed that depending on the $\omega_f/\tau_f$ ratio several stochastic generation regimes interspersed with regular generation regimes can be achieved in an AFM SHO that can be used in spintronic sources of random signals and various nano-scale devices utilizing random signals including the spintronic p-bit device considered in this paper.
The obtained results are important for the development and optimization of spintronic devices able to generate and process (sub-)THz-frequency random signals promising for ultra-fast probabilistic computing, cryptography, secure communications, etc.
\end{abstract}

\maketitle


\section{Introduction}

Antiferromagnetic (AFM) spintronics is an emergent field of science and technology focused on the study and utilization of static and dynamical magnetic states in antiferromagnets (AFMs) for the development and creation of various spintronic devices \cite{AFMsp16,AFMsp18PLA,AFMsp18NatPhys,AFMsp18RMP,AFMsp20,AFMsp23}.
Although the evolution of AFM spintronics in the present time generally seems similar to the history of conventional ferromagnetic spintronics, the development, and implementation of AFM devices turned out to be a more complex task than that for conventional spintronic devices based on ferromagnets \cite{AFMsp16,AFMsp18PLA,AFMsp18NatPhys,AFMsp18RMP,AFMsp20,AFMsp23,
Gomonay10,Gomonay14,Cheng16,Khymyn,Gen17PRAppl,Mem17,Neuro18SciRep,Neuro18JAP,ATJ18,
AFMOsc19,AFM19,AFMOsc20,ATJdetector20,Neuro20,AFMDev20,Ovcharov22}.
Nevertheless, ``pure AFM'' devices consisting AFM layers only can demonstrate ultra-fast magnetization dynamics and state switching with characteristic frequencies lying in a (sub-)THz band in the absence of a bias DC magnetic field \cite{AFMsp16,AFMsp18PLA,AFMsp18NatPhys,AFMsp18RMP,AFMsp20,AFMsp23,Gomonay14,Cheng16,Khymyn,
Gen17PRAppl,Mem17,Neuro18SciRep,Neuro18JAP,ATJ18,AFMOsc19,AFM19,AFMOsc20,ATJdetector20,
Rnd13,Rnd21,Neuro20,AFMDev20,Ovcharov22}, while spintronic devices based on ferromagnets usually require a bias DC magnetic field to gain high enough operation frequencies; for the reasonable bias fields, these operation frequencies usually does not exceed $\sim 50-80$~GHz for devices with ferromagnetic layers only or combined devices consisting both ferromagnetic and AFM layers \cite{FMsp04,FMspBook12,FMsp11,FMsp15,Gen}.
Due to that fact, ``pure AFM'' devices have a great potential for application in the field of terahertz science and technology (security systems, medical imaging, communications, material physics, etc.), as well as in ultra-fast conventional and neuromorphic electronics including magnetic memory, artificial intelligence systems, etc. \cite{AFMsp16,AFMsp18PLA,AFMsp18NatPhys,AFMsp18RMP,AFMsp20,AFMsp23,
Gomonay10,Gomonay14,Cheng16,Khymyn,Gen17PRAppl,Mem17,Neuro18SciRep,Neuro18JAP,ATJ18,
AFMOsc19,AFM19,AFMOsc20,ATJdetector20,Rnd13,Rnd21,Neuro20,AFMDev20,Ovcharov22}

One of the base key elements of AFM spintronics is an AFM spin Hall oscillator (SHO) \cite{AFMsp16,AFMsp18PLA,AFMsp18NatPhys,AFMsp18RMP,AFMsp20,AFMsp23,
Cheng16,Khymyn,Gen17PRAppl,Neuro18SciRep,ATJ18,AFMOsc19,AFMOsc20,Rnd13,Rnd21,Neuro20,AFMDev20,Ovcharov22}, which plays the same role as the spin-transfer-torque oscillator or ferromagnetic spin Hall oscillators in conventional (ferromagnetic) spintronics \cite{FMspBook12,Gen,FMsp16}.
Although a ``pure AFM'' SHO has been not experimentally studied yet (mainly due to the limitations in the fabrication technology of high-quality AFM samples), a lot of theoretical proposals for such devices have been made over the last years \cite{Cheng16,Khymyn,Gen17PRAppl,Neuro18SciRep,ATJ18,AFMOsc19,AFMOsc20,Rnd21,Ovcharov22} after early pioneer works \cite{Gomonay10,Gomonay14} by Gomonay and Loktev.
The mentioned papers \cite{Cheng16,Khymyn,Gen17PRAppl,Neuro18SciRep,ATJ18,AFMOsc19,AFMOsc20,Ovcharov22} mainly analyze possible schematics of bilayer or multilayer AFM SHOs and the principles of extraction of the generated AC signal power from these nanostructures (via the spin pumping and the inverse spin Hall effect \cite{Cheng16,Khymyn,Neuro18SciRep}, the direct emission of magnetodipolar radiation \cite{Gen17PRAppl}, etc.), while the question about the shape of the amplitude-frequency curve, statistical and spectral properties of the generated output AC signals in SHOs remains almost undiscussed (see  \cite{Rnd13,Rnd21}; this question was also partially studied for AFM SHO-based detectors in \cite{ATJdetector20}).
Nowadays many efforts in the field of AFM spintronics are concentrated on the development of coherent AFM devices \cite{Gen17PRAppl,AFMsp23}, however, one should note that \emph{random} or \emph{pseudo-random} AC signals generated in an AFM SHO can also play an important role in many promising applications including ultra-fast probabilistic computing, cryptography, secure and military communications, etc. \cite{AFMsp16,AFMsp18PLA,AFMsp18NatPhys,AFMsp18RMP,AFMsp20,Rnd13,Rnd21}

In this paper, we consider the stochastic magnetization dynamics in an AC current-driven Josephson-like bilayer AFM SHO consisting of a heavy metal layer and a NiO AFM layer having a biaxial magnetic anisotropy \cite{Khymyn} (Fig. \ref{I1}).
Previously it was shown that such an AFM SHO biased both by a DC and a time-harmonic AC currents can perform as a quasi-coherent AC signal source \cite{Khymyn} or can work as a comb generator (artificial ``neuron'') \cite{Neuro18SciRep}, and it can be the base element for various neuromorphic logic circuits \cite{Neuro18JAP}.
In addition to that it was demonstrated that such a DC current-biased SHO can be synchronized to a weak external AC signal demonstrating conventional \cite{Sl1} or fractional \cite{Sl2} synchronization.
In contrast to that below we consider the case when \emph{no bias DC current} is subjected to an SHO, thus, its behavior depends on the applied \emph{pure} time-harmonic AC current only.
Our model (presented in Sec.~\ref{S@Model}) also differs from the model studied in  \cite{Rnd21}, where the chaotic dynamics in an AFM SHO was analyzed in the parametric space (DC bias magnetic field, DC bias current).
In our case, using numerical calculations, we show that depending on the amplitude and frequency of the supplied AC current, several regimes of stochastic and quasi-coherent generation can be excited.
We analyze the statistics of the generated pseudo-random signal and propose the schematics of a p-bit spintronic device, which utilizes one of the revealed stochastic generation regimes of an SHO.
We believe that our results can be important for the development and optimization of spintronic devices operating with (sub-)THz-frequency random AC signals, which can be useful for some challenging applications.


\section{Model}
\label{S@Model}

\begin{figure}
 \includegraphics[width=1.0\columnwidth]{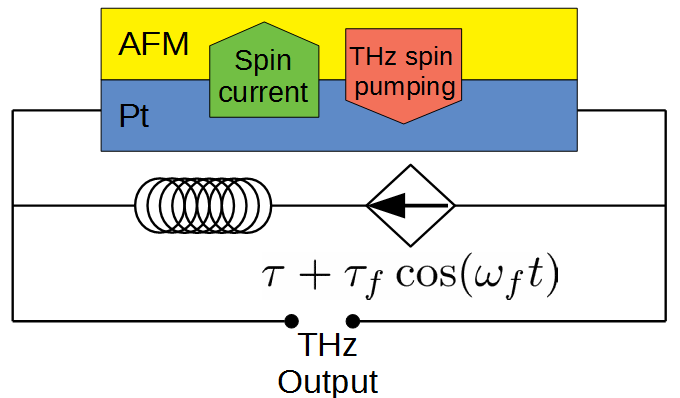}\\
\caption{(Color online) Schematics of the considered SHO having a thin AFM layer (NiO) covered by a layer of a heavy metal (Pt). Current density $\tau + \tau_f \cos\omega_f t$ is subjected to the Pt layer ($\tau = 0$ in this paper). The output AC signal is received due to the inverse spin Hall effect \cite{SHE15} at the AFM/Pt interface.}
\label{i1}
\end{figure}

The considered antiferromagnetic SHO consists of two layers: a heavy metal (platinum, Pt) layer characterized by a strong spin-orbit coupling and an antiferromagnetic (NiO) layer (Fig. \ref{i1}).
Electric DC current subjected to the Pt layer creates a transverse spin current flowing into the adjacent AFM layer due to the spin Hall effect \cite{SHE15}.
The action of spin current manifests itself like the spin-transfer torque \cite{STT18} resulting in the excitation of THz-frequency rotation of magnetic sublattices in the easy plane of an AFM.
On the other hand, the rotation of these magnetic sublattices creates an inverse spin current that flows back into a heavy metal layer via the AFM/metal interface and then converts to an electric current due to the inverse spin Hall effect \cite{SHE15}.
It was shown in  \cite{Khymyn} that this output current has only a DC component if magnetic sublattices of an AFM layer rotate at a constant speed.
But when the AFM layer of an SHO has a rather strong in-plane anisotropy in the easy plane, the rotation of magnetic sublattices of an AFM layer appears to be nonuniform (rotation speed changes) in time and the output electric current acquires a THz-frequency component.
This THz-frequency electric current injected in the Pt layer of the considered NiO/Pt SHO structure manifests itself as the device's output AC signal.
Thus, by applying a DC current density to the system, one can obtain a THz-frequency output electromagnetic signal, i.e. the device works as a tunable current-driven THz-frequency signal source  \cite{Khymyn,Neuro18SciRep,Neuro18JAP}.
Furthermore, by adding an AC signal to the DC component of the current density, one can further manipulate the system dynamics.

Magnetization dynamics in an AFM layer of the considered SHO can be described by an equation, which is derived from the so-called $\sigma$-model and has the form \cite{Khymyn}:
\begin{equation}
\frac{1}{\omega_{ex}}\ddot\phi+\alpha \dot\phi+\frac{\omega_e}{2}\sin 2\phi=\tau+\tau_f \cos\omega_f t \, .
\label{init}
\end{equation}
It describes the AFM dynamics in terms of a single parameter -- the Neel vector azimuthal angle $\phi \equiv \phi(t)$; $\dot\phi \equiv d\phi/dt$, $\ddot\phi \equiv d^2\phi/dt^2$.
In the right-hand side of Eq. (\ref{init}) we have a driving force, acting on the oscillator in the form of applied DC ($\tau$) and AC ($\tau_f \cos\omega_f t$) current density written in the units of angular frequency; $\tau_f$ and $\omega_f$ are the amplitude and angular frequency of the AC current density, respectively.
In Eq. (\ref{init}) $\omega_{ex}=\gamma H_{ex}$ is the exchange angular frequency, which depends on the exchange field $H_{ex}$ in an AFM, $\omega_e = \gamma H_e$ is the easy axis angular frequency that is proportional to the easy axis anisotropy field $H_e$, $\alpha$ is the damping parameter, and $\gamma$ is the modulus of the gyromagnetic ratio.
The $\sigma$-model approximation remains valid as long as the dynamical ground state of the system corresponds to the Neel vector rotation in the easy plane \cite{Khymyn}, which is fulfilled for the case considered in this paper.

Dynamics of Eq. (\ref{init}) was previously analyzed in \cite{Khymyn, Sl1, Sl2}, so we describe just some of its key features that could be important for further analysis.
Eq. \ref{init} describes a nonlinear damped oscillator driven by an external force.
Increase of the driving force $\tau$ leads to the excitation of rapid oscillations of the Neel vector angle $\phi$ with the rotation speed $\dot\phi$.
This generation has a threshold and appears for $\tau > \tau_{th}=\omega_e/2$ only (see \cite{Khymyn}).
It was shown in  \cite{Khymyn} that the Neel vector rotation speed or the angular frequency of oscillations, $\dot\phi$, can be estimated as $\omega_{gen}=2\tau / \alpha$ for $\tau \gg \tau_{th}$.
The angle rotation speed $\dot\phi$ is an important parameter, because the output AC electric field $E$ and, thus, the output AC current density $j_{out}$, produced via the inverse spin Hall effect \cite{SHE15} is directly proportional to it  \cite{Khymyn}: $j_{out} \sim E \sim \dot\phi$.

Eq. \ref{init} can be written in the dimensionless form: $\xi\ddot\phi+\dot\phi+\sin\phi=i_0+i_1 \cos\Omega t$,
where $\xi=\omega_e / \omega_{ex} \alpha^2$, $\Omega= \alpha \omega_f / \omega_e$, $i_0=2 \tau / \omega_e$,  $i_1=2 \tau_f / \omega_e$.
This form of the equation is similar to the one describing the dynamics in Josephson junctions  \cite{Okt1,Okt2,Kau1,Kau2}.
It is well known \cite{Pykovskiy}, that solution of such an equation can exhibit a stochastic behavior in the presence of an external AC signal due to a strong nonlinearity.
Moreover, parameter $\xi$ describes the chaos ``strength'' -- the larger it is, the more prominent stochastic dynamics can be observed \cite{Kau1}.
Naturally at $\xi\rightarrow 0$ the system does not demonstrate any stochastic behavior, because \ref{init} becomes a first-order differential equation that is not sufficient to support stochastic dynamics \cite{Pykovskiy}.

\begin{figure*}
 \includegraphics[width=1.9\columnwidth]{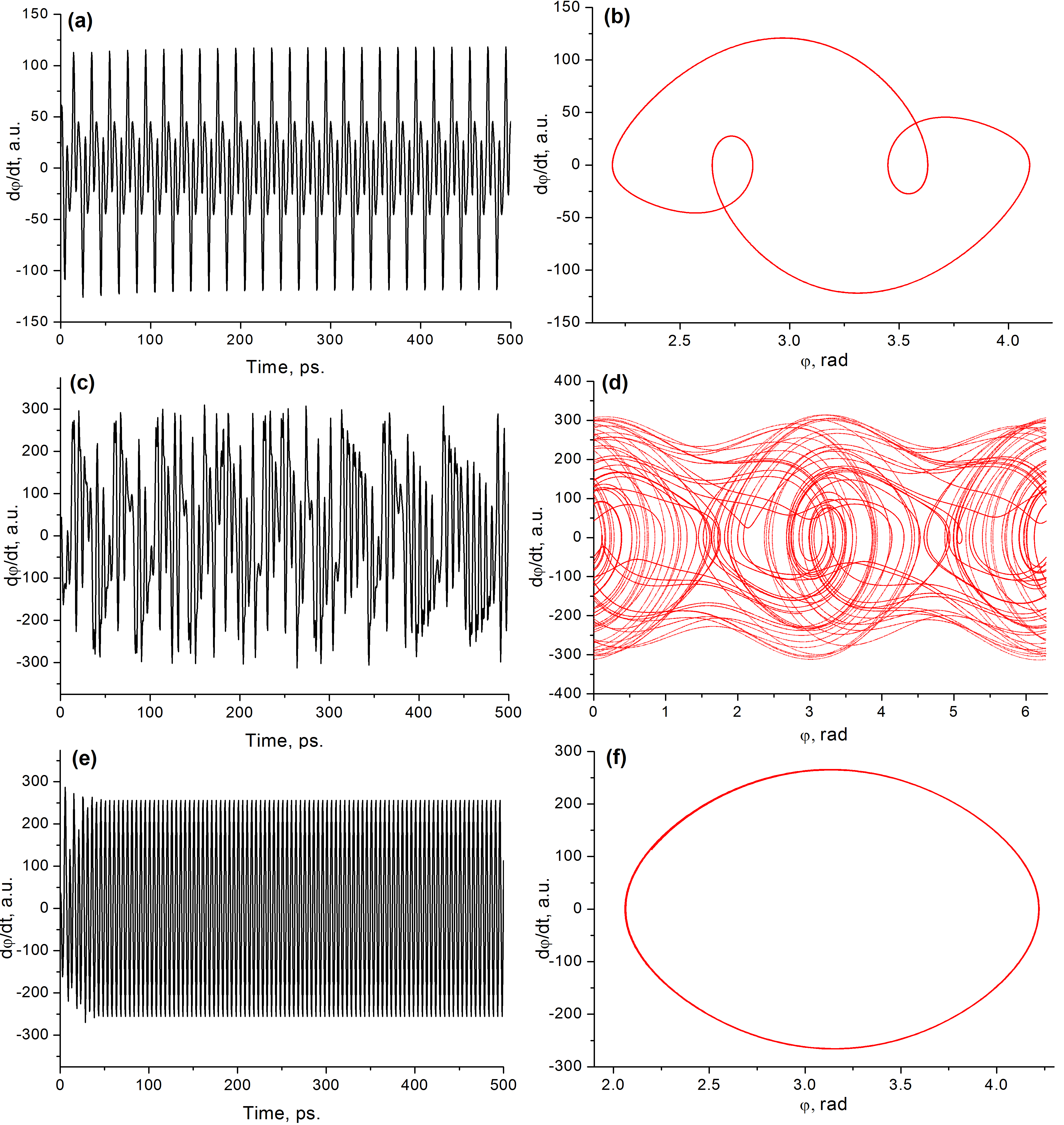}\\
\caption{(Color online) Time dependence of the Neel vector rotation speed $\dot\phi$ (left, black line) and the corresponding phase portrait of the system in ($\phi,\dot\phi$) coordinates (right, red line) for the fixed AC signal amplitude $\tau_f/\omega_e=0.5$ and various AC signal frequency $\omega_f$: (a)-(b) $\omega_f=2 \pi \cdot 50$ GHz; (c)-(d) $\omega_f=2 \pi \cdot 150$ GHz; (e)-(f) $\omega_f=2 \pi \cdot 200$ GHz. All curves are calculated numerically from Eq. (\ref{init}). All other parameters are indicated at the beginning of Sec.~\ref{SS@Params}.}
\label{f1}
\end{figure*}

In the following, we consider the case, when a pure AC signal is subjected to a Josephson-like AFM SHO \cite{Khymyn}, while the DC component of the bias current is \emph{absent}.
Thus, we have the case $\tau=0$, $\tau_f \neq 0$ in Eq. (\ref{init}).
Note that many features of an AFM SHO biased by a DC and a weak AC signal including synchronization and fractional synchronization were studied before in \cite{Khymyn, Neuro18SciRep, Neuro18JAP, Sl0,Sl1,Sl2}. However, in this paper, we analyze a significantly different case, when the bias DC current density $\tau$ is \emph{zero}, and an oscillator is biased by the pure AC current only.
Thus, under any circumstance, the current density $\tau_f \cos\omega_f t$ could not be considered as a small quantity, which was done before in \cite{Sl0,Sl1,Sl2}.
Taking this into account, further we will search the conditions in the coordinates ($\tau_f, \omega_f$) needed for the formation of dynamical chaos in the considered system, and investigate the properties of excited chaotic dynamics in an AFM SHO.

To analyze the dynamics of the system we numerically solve Eq. (\ref{init1}) and obtain the dependencies $\phi \equiv \phi(t)$ and $\dot\phi \equiv \dot\phi(t)$.
We also use the phase portrait method introduced for AFM SHOs in \cite{Sl0}.
In the case of Eq. (\ref{init}) the phase portrait is a two-dimensional surface in the coordinates ($\phi, \dot\phi$) (see Fig. \ref{f1}).
Due to the cyclic condition, where the angles $\phi = 0$ and $\phi = 2 \pi$ correspond to the same system's state, the phase space of the system, in fact, is a surface of a cylinder.
Finally, to exclude from the analysis unstable magnetization dynamics caused by the transition processes at the early stages of the system evolution, we omit the numerically calculated data $\phi, \dot\phi$ for the initial 100~ps when plotting the phase portraits.


\section{Results and discussion}
\label{S@Results}

\subsection{Phase portraits of the system}
\label{SS@Params}

In this part of the paper, using the phase portrait method, we investigate the regimes of the AFM SHO operation in the absence of an external DC bias ($\tau = 0$ in Eq. (\ref{init})) and determine the control parameters $\tau_f$ and $\omega_f$ of an input AC signal, when the system behavior becomes chaotic and the SHO goes into the state of dynamical chaos.
In that state output signal generated by the AFM SHO ($\sim \dot\phi$) becomes (pseudo-)random, which can be used in some applications.

For a quantitative description of the magnetization dynamics in an AFM SHO, we use the following typical parameters \cite{Khymyn}: easy plane field $H_e=625$~Oe, exchange field $H_{ex}=9\cdot10^6$~Oe, damping parameter $\alpha=2\cdot10^{-3}$.
These parameters give the value of the chaos `` strength'' $\xi \simeq 17$.
It should be noted, that in works regarding Josephson junction dynamics \cite{Okt1,Okt2,Kau1,Kau2} characteristic values of $\xi$ were about several hundred, therefore, in the case of an AFM SHO with chosen parameters one can expect that the stochastic features of the system will be less prominent than in the case of Josephson junctions \cite{Okt1,Okt2,Kau1,Kau2}.

From the numerical solution of Eq. (\ref{init}) one can obtain typical dependencies of $\dot\phi(t)$ and phase portraits in coordinates ($\phi(t)$, $\dot\phi(t)$) for the AFM SHO under the action of time-harmonic AC current with fixed amplitude $\tau_f = 0.5 \omega_e$, but different angular frequencies $\omega_f$ (Fig.~\ref{f1}).
As one can see from Fig.~\ref{f1}(a) the dynamics of the system is described by relatively small oscillations near the equilibrium $\dot\phi=0$ and a set of periodic large pulses, when the AC current frequency is $\omega_f = 2\pi \cdot 50$~GHz.
This behavior corresponds to the phase portrait ($\phi(t)$, $\dot\phi(t)$) in the form of a complex double closed loop (Fig. \ref{f1}(b)).
Bigger loops correspond to a periodic spike-like pulse generation, while smaller loops correspond to the case of small oscillations, which take place between the strong pulses (see Fig. \ref{f1}(a,b)).

When AC signal frequency $\omega_f$ is increased up to $\omega_f = 2\pi \cdot 150$~GHz and at the same time the current amplitude remains constant, the system dynamics suffer drastic changes (Fig. \ref{f1}(c,d)).
As it can be seen from Fig. \ref{f1}(c), under the action of a 150~GHz AC signal the system dynamics becomes stochastic: one can observe no (quasi-)periodic pulse generation.
Instead of that we have a (pseudo-)random output AC signal having approximately constant frequency, but time-dependent \emph{(pseudo-)random} amplitude.
According to that, such a signal can be treated as a (quasi-)periodic signal with random amplitude modulation.
Phase portrait of this stochastic signal generated in the AFM SHO is far more complicated than in the case of regular SHO generation regimes (compare Fig. \ref{f1}(d) with Fig. \ref{f1}(b, f)).
No longer it consists of rather clear closed loop(s) in some relatively small confined region in the parametric space ($\phi$, $\dot\phi$).
Instead of that the phase portrait occupies the entire plane ($\phi$, $\dot\phi$).
This regime is one of the stochastic regimes of AFM SHO operation that will be investigated below.

Finally, by further increasing AC signal frequency, one can achieve that at sufficiently higher frequencies, for instance, $\omega_f = 2\pi \cdot 200$~GHz, the system can demonstrate the regular (deterministic) dynamics again (Fig. \ref{f1}(e, f)).
Furthermore, in that case, the system dynamics is even simpler than in the case of smaller frequency $\omega_f = 2\pi \cdot 50$~GHz (see Fig. \ref{f1}(a, b)), because $\dot\phi(t)$ has a form of regular periodic oscillations around the ground state $\dot\phi=0$.
Thus, the phase portrait of the system driven by an AC current with a rather high frequency is the simplest of all the three considered cases and contains only one ``almost perfect'' closed loop (see Fig. \ref{f1}(f)).

\begin{figure}
 \includegraphics[width=1.0\columnwidth]{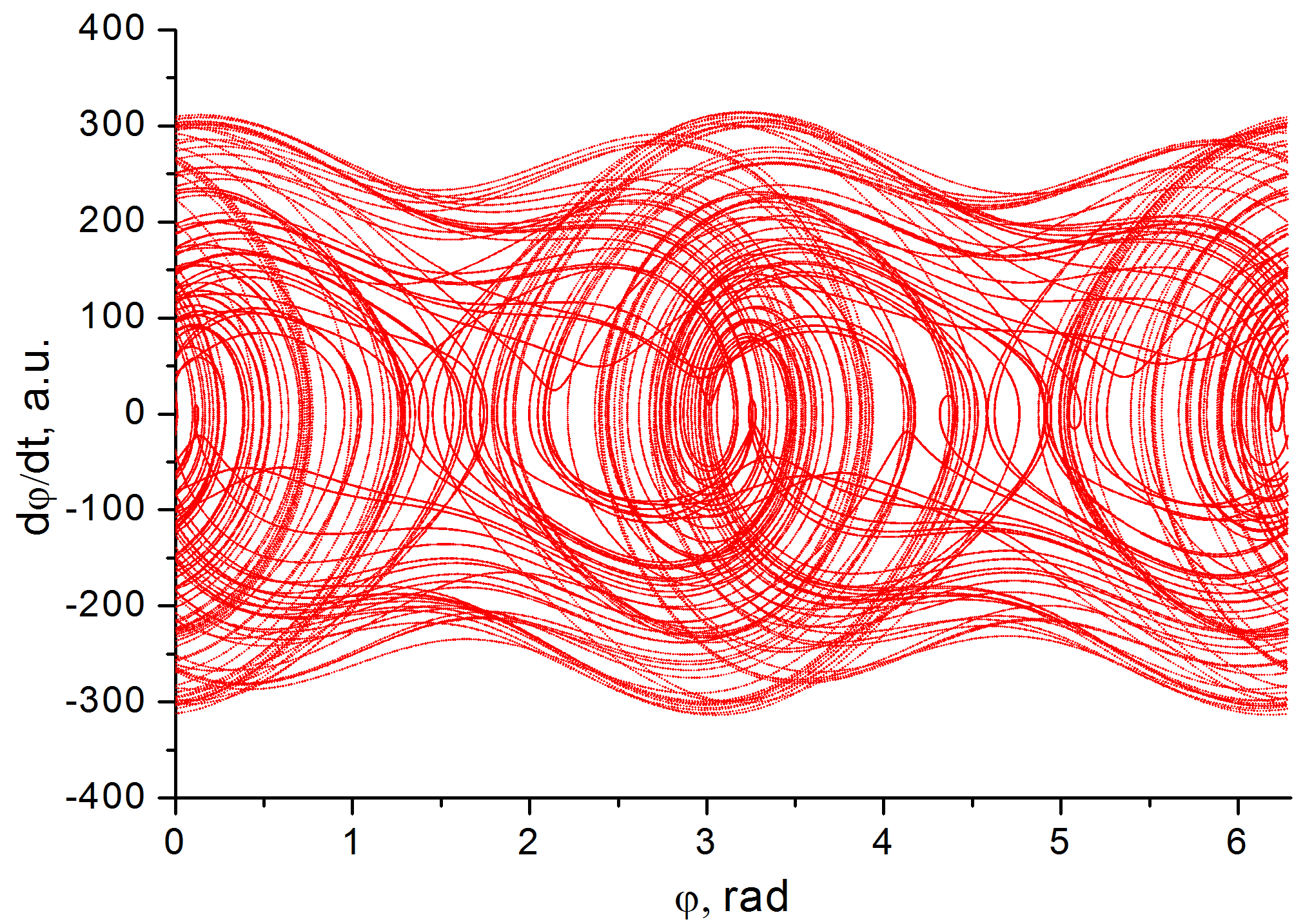}\\
\caption{(Color online) Numerically calculated phase portrait of the system for a long period of time $T_{max}=2000 ps$. AC signal parameters: $\tau_f/\omega_e=0.5$, $\omega_f=2 \pi \cdot 150$ GHz. All other parameters are the same as indicated at the beginning of Sec.~\ref{SS@Params}.}
\label{f2}
\end{figure}

To illustrate, how the phase portrait of the system evolves in the stochastic regime, when the simulation time increases, we calculate two phase portraits for different simulation times $T_{max}$: $T_{max} = 500$~ps (Fig. \ref{f1}(d)) and $T_{max} = 2000$~ps (Fig.\ref{f2}).
As one can see, with an increase of the simulation time $T_{max}$, the phase trajectories cover more and more closely the phase space ($\phi,\dot\phi$).
Such behavior indicates the stochastic nature of the system dynamics  \cite{Pykovskiy}.
Also, the ``density'' of calculated phase trajectories in the phase space is not uniform.
It follows from Fig.~\ref{f2} that there are several phase space regions with larger phase trajectory ``density'', which could correspond to some quasi-stable states of the AFM SHO. From that point of view, the nature of the observed stochastic generation regime could be explained by the random hopping of the SHO's work point between several quasi-stable states under the action of applied AC current, which require an additional study.

\subsection{Region of chaos existence and the output signal statistical properties}

To investigate the system behavior and find the region of parametric space ($\tau_f$, $\omega_f$) corresponding to the chaotic dynamics of an AFM SHO we calculate the Lyapunov exponents that determine the type of system dynamics \cite{Lyap,Rnd21}.
These exponents characterize the instability of the system phase trajectories with respect to the initial conditions.
A positive Lyapunov exponent means that the phase trajectories diverge indicating the chaotic system dynamics.
In the case of Eq. (\ref{init}), the system state can be described by a two-dimensional vector in the phase space ($\phi,\dot\phi$), thus, we need to calculate two Lyapunov exponents, $L_1$ and $L_2$, using the standard procedure \cite{Pykovskiy, Lyap, Rnd21}.
We calculate these exponents in a wide range of control parameters ($\tau_f$, $\omega_f$) to determine regions, where one of the exponents becomes strictly positive, $L_j>0$.
This will indicate the existence of the state of chaotic dynamics for the chosen control parameters.

\begin{figure}
 \includegraphics[width=1.0\columnwidth]{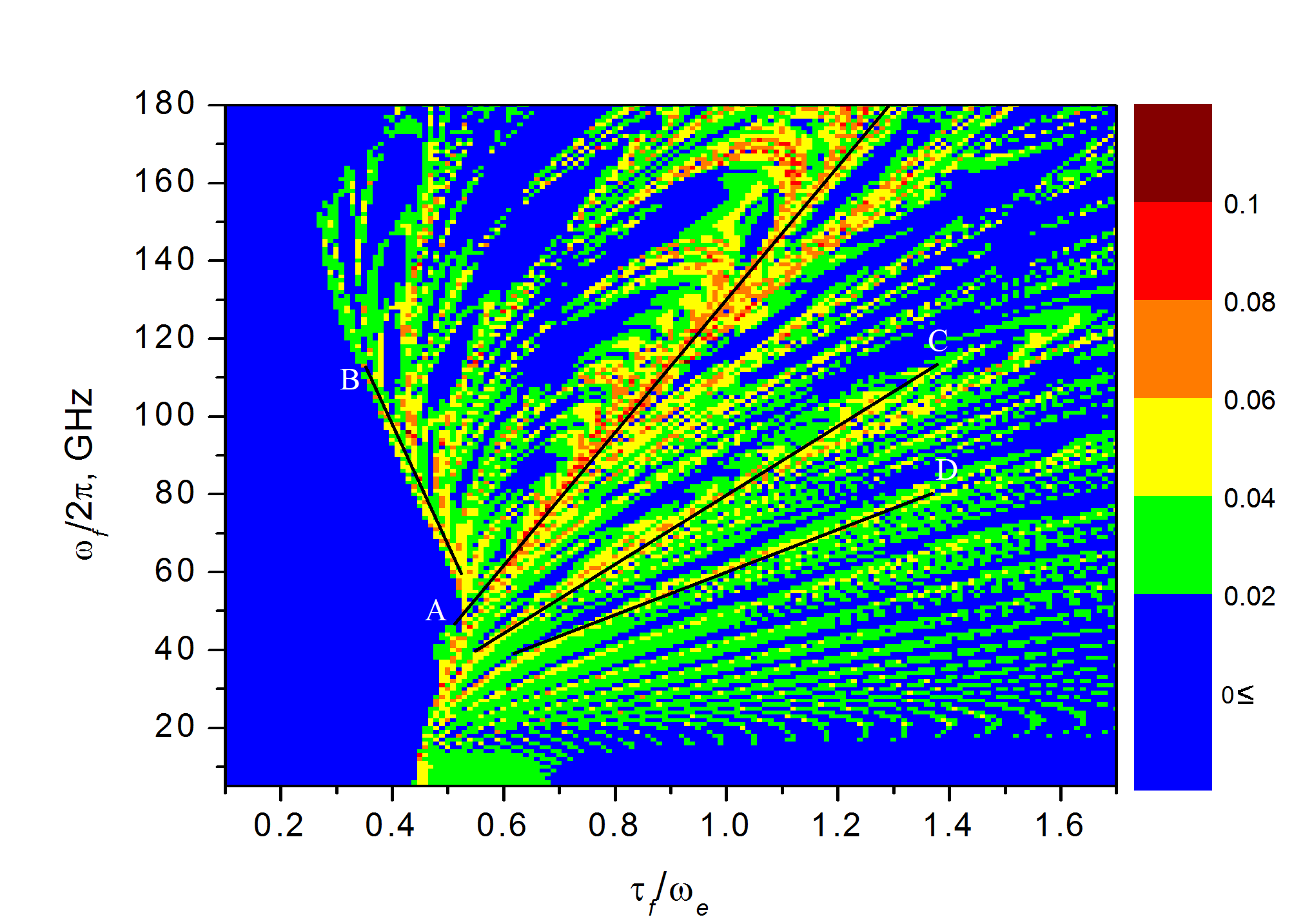}\\
\caption{(Color online) Map of the largest Lyapunov exponent calculated using the standard procedure \cite{Pykovskiy, Lyap, Rnd21} from Eq.~(\ref{init}). Straight black lines labeled A, B, C, D correspond to different stochastic generation regimes of the considered AFM SHO. Simulation parameters are indicated in Sec.~\ref{SS@Params}.}
\label{L1}
\end{figure}

Calculated values of the largest Lyapunov exponent depending on parameters ($\tau_f$, $\omega_f$) are shown in Fig.~\ref{L1}.
As we can see, this map is rather complicated.
There is no chaotic behavior for $\tau_f/\omega_e<0.3$ for any signal frequency up to $\omega_f/2\pi = 180$~GHz.
At larger signal amplitudes, when $\tau_f/\omega_e \ge 0.3$, the regions of chaotic dynamics are interspersed by the narrow regions of regular dynamics with $L \leq 0$.
Generally, the regions of chaotic and regular dynamics form a rather fuzzy ``striped'' pattern, where each ``strip'' corresponds to one of two possible regimes (chaotic or regular).

Due to the complexity of the calculated map for an AFM SHO operation regimes (Fig.~\ref{L1}), the exact determination of the region boundaries for the chaotic dynamics is a rather complicated task, that lies beyond the scope of this work (an example of such an analysis for a simpler model of a DC-biased AFM SHO is presented in \cite{Rnd21}).
However, from Fig.~\ref{L1} one can make two basic conclusions.
i) It is clear that chaotic dynamics of the system (\ref{init}) exists in a rather wide region of parametric space ($\tau_f$, $\omega_f$) and, thus, it can be easily tuned by applying a driving AC current with some particular amplitude $\tau_f$ and/or frequency $\omega_f$.
ii) For the AFM SHO with chosen parameters operating at rather low input signal frequencies $\omega_f$, the regime of stochastic generation is most clearly observed for approximately linear relation between the AC signal frequency $\omega_f$ and the signal amplitude $\tau_f$.
This feature, based on the calculation of $\omega_f/\tau_f$ ratio, probably could be used for the classification of various stochastic generation regimes.
For instance, using the proposed technique for Fig.~\ref{L1}, one can mark out one clearly seen stochastic generation regime A with $\omega_f/\tau_f \approx 107.5$ and several other regimes of chaotic dynamics: B ($\omega_f/\tau_f \approx -178$), C ($\omega_f/\tau_f \approx 47$), and D ($\omega_f/\tau_f = 31.5$) (see straight black lines labeled A, B, C, D in Fig.~\ref{L1}).

Output AC signal of an SHO operating in the stochastic generation regime keeps its periodic nature with the oscillation frequency that is close to the driving frequency $\omega_f$, while the amplitude of the generated signal changes (pseudo-)randomly in time (Fig. 2(c)).
Moreover, an entire AFM SHO in the regime of stochastic generation can be considered as a random amplitude modulator: when it receives an input time-harmonic signal of frequency $\omega_f$ and constant amplitude $\tau_f$, it outputs a periodic signal with the same frequency but (pseudo-)randomly modulated amplitude.
In the next section of the paper we will propose how this feature of the AFM SHO dynamics could be used in practice.

\begin{figure*}
 \includegraphics[width=1.9\columnwidth]{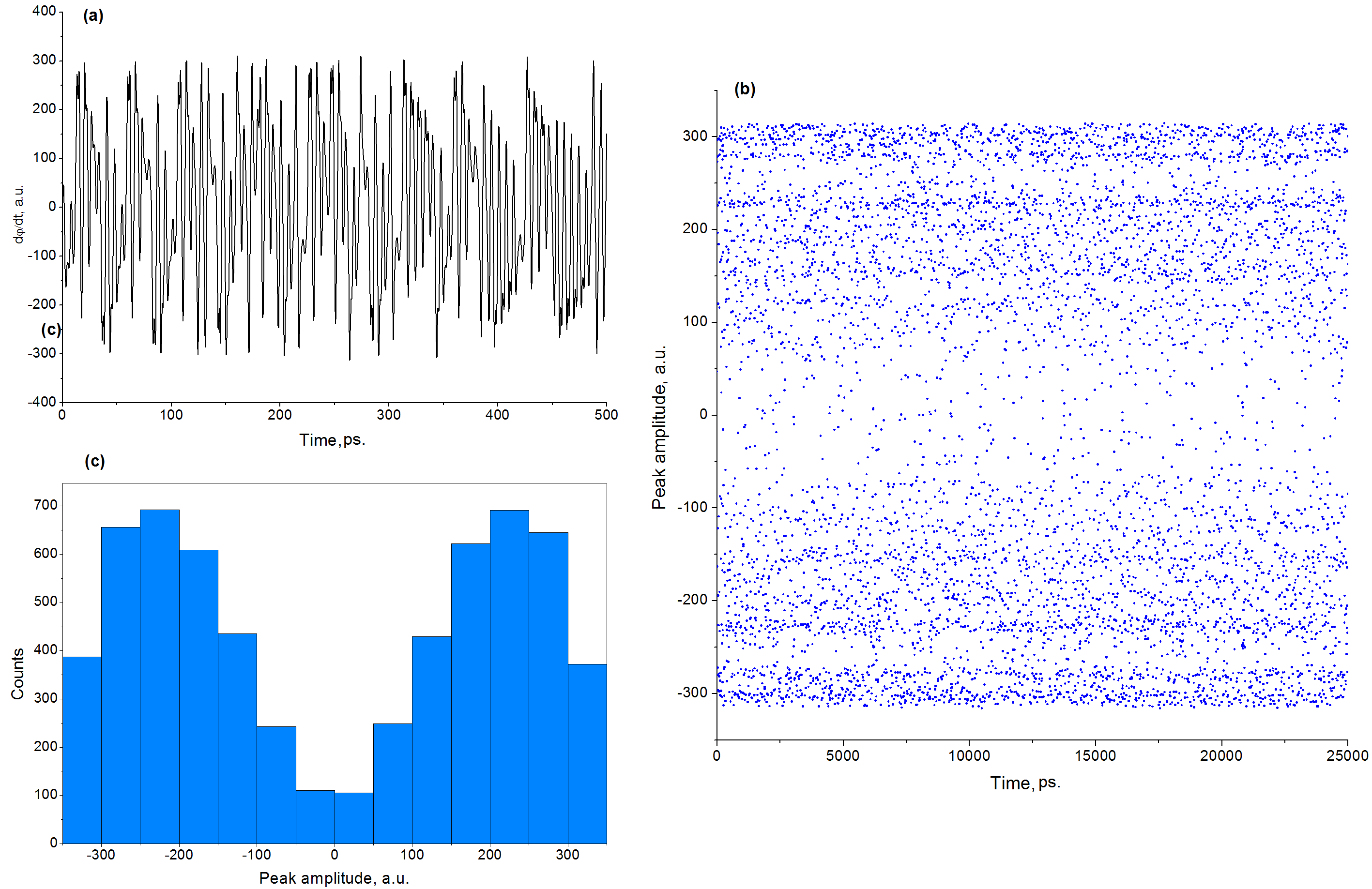}\\
\caption{(Color online) Statistics of the generated stochastic output signal of a Josephson-like AFM SHO: (a) typical time profile of the generated signal; (b) signal peak amplitudes as a function of the simulation time; (c) histogram of the peak amplitudes. AC signal parameters: $\tau_f/\omega_e=0.5$, $\omega_f=2 \pi \cdot 150$ GHz. All other parameters are the same as indicated at the beginning of Sec.~\ref{SS@Params}.}
\label{S1}
\end{figure*}

Statistical parameters of the typical stochastic output AC signal generated by the AFM SHO are shown in Fig. \ref{S1}.
The figure includes three panels illustrating the time dependence of the generated signal (a), the distribution of signal peak amplitudes in time for a rather long period of time (b), and the histogram of signal peak amplitudes distribution (c).
Both last figures have calculated for the simulation time increased by a factor of 50, up to $T_{max} = 25$~ns, in order to collect enough data points for adequate statistical analysis.

As one can see, the resulting histogram of the signal peak amplitudes distribution is rather nontrivial (see Fig. \ref{S1}(c)).
It is almost symmetrical for positive and negative amplitudes, however, the distribution itself deviates greatly from the classic Poisson distribution.
By comparing these results with the available data for superparamagnetic tunnel junctions \cite{Maj, SPTJ20}, where hopping between the states follows the Poisson statistics, one can conclude that the complicated nature of the output SHO signal statistics may lead to some difficulties in practical applications of the AFM SHO operating in a stochastic operation regime.
However, there are some ways to overcome those complications, one of which will be discussed in the next section.

\section{Possible realization of the p-bit using regime of stochastic generation of an AFM SHO}

The so-called p-bit is the main building block of ``probabilistic spin logic'' \cite{Neuro20,Camsari,NeuroEl}.
The output signal of the p-bit can be written as:
\begin{equation}
m(t)= \sgn[ \rand(-1,1) +\tanh (I(t))] \, ,
\label{pb1}
\end{equation}
where $\rand(-1,1)$ is the random number, uniformly distributed in $(-1,1)$ interval, $\sgn$ is the signum function, $I(t)$ is an external input signal, which allows one to shift p-bit state (its output) from the state with equal probabilities for ``1'' or ``$-1$'' output.
The physical implementation of such p-bits can be based on utilization of stochastic nanomagnets with low-energy barriers  \cite{Neuro20,Camsari,NeuroEl}.

\begin{figure}
 \includegraphics[width=1.0\columnwidth]{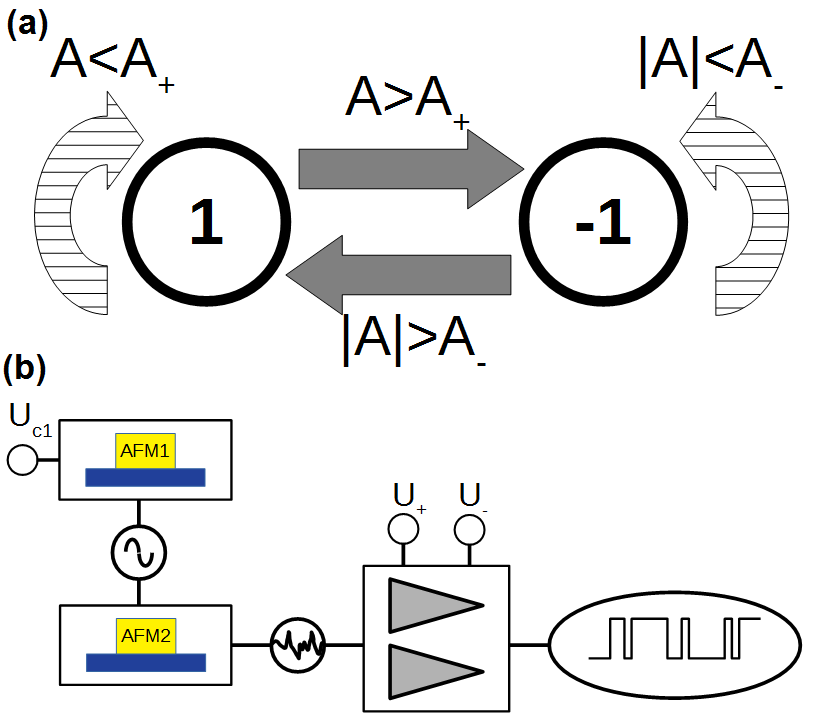}\\
\caption{(Color online) The Markov-like stochastic signal converter: (a) schematic illustration of the device operation scheme; (b) principal scheme for the p-bit implementation using an AFM SHO.}
\label{P1}
\end{figure}

In the case of an AFM SHO, the task of constructing a p-bit is rather complicated.
Analyzing the evolution of stochastic signal in time (see Fig. 2(c)), it is difficult to say, how a system with the characteristic equation similar to (\ref{pb1}) could be implemented.
Moreover, as it was shown in the previous section, signal peak distribution greatly deviates from Poisson distribution (see Fig. \ref{S1}(c)), which prevents the use of traditional approach \cite{Neuro20,Camsari,NeuroEl} for p-bit implementation.

To solve this problem and convert the stochastic output AFM SHO signal (Fig. 2(c)) to the typical p-bit output signal with controlled state parameters, we propose the following scheme.
We consider the stochastic signal from the SHO as some controlling signal for a Markov-like system, which has two states ``1'' and ``$-1$'' with deterministic rules of switching between the states as it is shown in Fig. \ref{P1}(a).
We introduce two controlling amplitudes: $A_{+}$ and $A_{-}$ that govern the system dynamics.
If peak amplitude does not exceed these values, the system remains in its present state.
However, if positive peak amplitude exceeds $A_{+}$ for the ``1'' state, then the system switches to the ``$-1$'' state.
The same happens if the peak amplitude exceeds $A_{-}$ amplitude: the system from the ``$-1$'' state will switch to the ``1'' state.
This behavior is schematically illustrated in Fig.~\ref{P1}(a).
By changing values of $A_{+}$ and $A_{-}$ one can change the probability of system being in the ``1'' or ``$-1$'' states.
Due to the symmetrical nature of the signal peak distribution, in the case of $A_{+}=A_{-}$, one should expect 50\%/50\% distribution of ``1'' and ``$-1$'' states. Shifting controlling amplitudes, for instance, $A_{+}>A_{-}$, should lead to the dominance of the ``$-1$'' state and vice-versa.

\begin{figure*}
 \includegraphics[width=1.9\columnwidth]{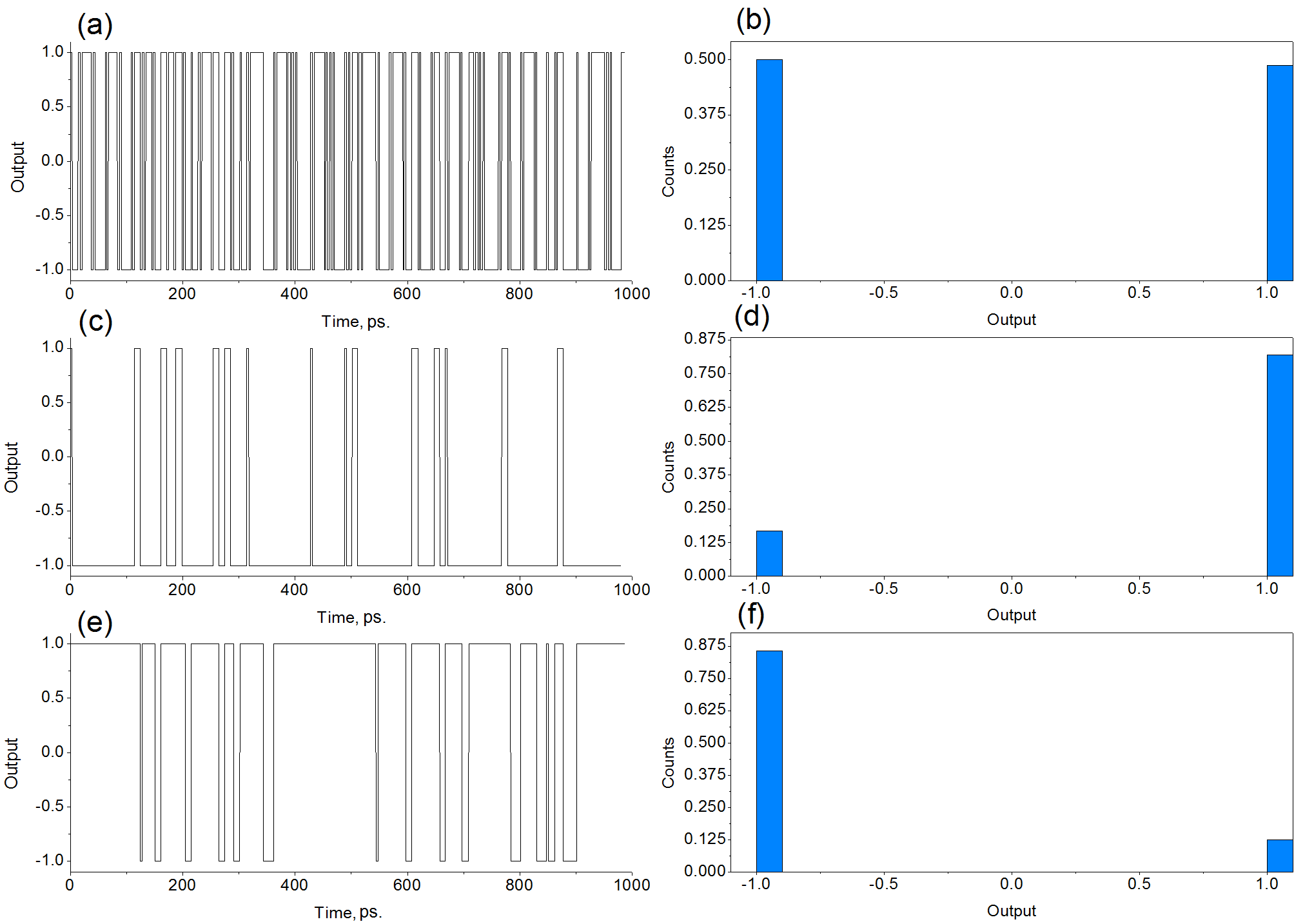}\\
\caption{(Color online) Output of the Markov-like signal convertor and corresponding system's states distribution for different values of contol amplitudes $A_{+}$ and $A_{-}$: (a)-(b) $A_{+}=A_{-}=150$; (c)-(d) $A_{+}=300, A_{-}=150$; (e)-(f) $A_{+}=150, A_{-}=300$. All other parameters are the same as indicated at the beginning of Sec.~\ref{SS@Params}.}
\label{P2}
\end{figure*}

Possible p-bit implementation using an AFM SHO is shown in Fig.~\ref{P1}(b).
It consists of two AFM SHOs and a controlling block with comparators that implement the scheme presented in Fig.~\ref{P1}(a).
The principle of operation of the scheme is as follows.
The first AFM SHO driven by a DC voltage $U_{c1}$ outputs a harmonic (sub-)THz signal via the inverse spin Hall effect \cite{SHE15}.
This signal is then applied to the second AFM SHO, which acts like a random amplitude modulator: while keeping the signal frequency stable, it randomly changes individual peak amplitudes, as shown above (see Sec.~\ref{S@Results}).
The random signal generated by the second SHO, which is similar to the signal shown in Fig.~\ref{f1}(c), acts as an input signal for the control block with two applied control voltages, $U_{+}$ and $U_{-}$.
It is transformed according to the scheme (Fig. \ref{P1}(a)) to the final output signal that looks like the traditional p-bit output signal.
Therefore, the final output signal simulates the switching between the ``1'' and ``$-1$'' states of a whole system with some particular probability defined by the ratio of voltages, $U_{+}/U_{-}$, applied to the comparator.

It should be noted, that the mathematical problem of construction of the principal probability function in a form $P_1(A_{+},A_{-},\omega_f,\tau_f)$ that returns the probability of the system being in the ``1'' state for given input signal parameters and controlling amplitudes lies far beyond the scope of this work.
However, we believe that such a function can be derived mathematically, and can be of great aid for future p-bit applications, allowing one to finely tune controlling amplitudes to achieve the desired p-bit output.

The proposed p-bit scheme utilizes two AFM SHOs for generating periodic and stochastic THz-frequency signals.
Moreover, using the neuromorphic computing approach one can utilize AFM SHOs for the third part of our scheme too \cite{Neuro18SciRep,Neuro18JAP,Neuro20}, realizing fast comparators based on AFM SHOs.
Thus, an entire p-bit can be fabricated in one technological process using AFM SHOs only, which is a great benefit of this scheme.
However, the main benefit of utilizing the AFM SHO is its ultra-fast switching from one state to another with a typical switching time of picoseconds, which is much faster compared to the systems based on superparamagnetic tunnel junctions \cite{Maj, SPTJ20}.

Finally, to illustrate how this Markov-like signal converter works, we apply to the device an input signal shown in Fig.~\ref{f1}(c) and consider the cases of different controlling amplitudes $A_{+}, A_{-}$.
The simulation results are shown in Fig.~\ref{P2}.
As expected, the change of control amplitudes $A_{+}, A_{-}$ substantially influences the ``preferred'' state of the system.

\section{Conclusions}

In this work, we numerically instigated the regimes of stochastic generation in a Josephson-like AFM SHO driven by a pure time-harmonic input AC current with some particular amplitude $\tau_f$ and angular frequency $\omega_f$.
We found that for an input signal having constant amplitude but variable frequency, depending on the frequency/amplitude ratio $\omega_f/\tau_f$ the oscillator can operate in various regimes, some of which demonstrate regular magnetic dynamics, while others reveal the chaotic dynamics.
The nature of the observed stochastic generation regimes probably could be explained by the random hopping of the oscillator state between the several quasi-stable regimes of regular magnetic dynamics under the action of applied AC current.
We numerically calculated a map of the SHO's Lyapunov exponents in the phase space ($\tau_f$, $\omega_f$), which displayed a rather fuzzy ``stripped'' pattern with interspersed regular and chaotic operation regimes, where the chaotic ones could be distinguished by the $\omega_f/\tau_f$ ratio.
The statistics of the stochastic signal generated in an AFM SHO demonstrated a substantial deviation from the Poisson distribution (typical for superparamagnetic tunnel junctions), which substantially complicates the creation of AFM SHO-based devices using conventional procedures for probabilistic computing. 
Nevertheless, the application of AFM SHOs for this task still could be possible if some unconventional procedures for probabilistic computing were used. 
To illustrate that we proposed a scheme of a p-bit device based on two AFM SHOs that allows one to overcome these difficulties and convert a stochastic output signal generated in an AFM SHO into a typical p-bit output signal corresponding to the switching between some two ``1'' and ``$-1$'' states. 
The obtained results are important for the development and optimization of AFM spintronic devices operating with (sub-)THz-frequency random signals including ultra-fast AFM SHO-based probabilistic computing, cryptography, and secure communications.

\section*{ACKNOWLEDGEMENTS}

This work was supported in part by grant No. 22BF07-03 from the Ministry of Education and Science of Ukraine, grant 16F-2022 from the National Academy of Sciences of Ukraine, the NATO SPS.MYP G5792 grant, and the STCU grant 9918 (IEEE Magnetism in Ukraine Initiative).

It is a pleasure to acknowledge many helpful discussions with Dr. A.V. Kovalenko (Taras Shevchenko National University of Kyiv).

The authors thank all brave defenders of Ukraine that made finalizing this publication possible.

\bibliography{export}

\end{document}